# The effect of anti-COVID-19 policies to the evolution of the disease: A complex network analysis to the successful case of Greece.

**Dimitrios Tsiotas[1,2*] and Lykourgos Magafas[3]**

[1]. Department of Regional and Economic Development, Agricultural University of Athens, Greece, Nea Poli, Amfissa, 33100, Greece.
[2]. Department of Planning and Regional Development, University of Thessaly, Pedion Areos, Volos, 38334, Greece.
Tel +302421074446, fax: +302421074493
[3]. Laboratory of Complex Systems, Department of Physics, International Hellenic University, Kavala Campus, St. Loukas, 65404, Greece.
E-mails: tsiotas@uth.gr – tsiotas@aua.gr; lmagafas@otenet.gr
[*]. *Corresponding author*

**Abstract**
Within the context that Greece promises a success story in the fight against the disease, this paper proposes a novel method to study the evolution of the Greek COVID-19 infection-curve in relation to the anti-COVID-19 policies applied to control the pandemic. Based on the ongoing spreading of COVID-19 and the insufficient data for applying classic time-series approaches, the analysis builds on the visibility graph algorithm to study the Greek COVID-19 infection-curve as a complex network. By using the modularity optimization algorithm, the generated visibility graph is divided into communities defining periods of different connectivity in the time-series body. These periods reveal a sequence of different typologies in the evolution of the disease, starting with a power pattern, where a second order polynomial (U-shaped) pattern intermediates, being followed by a couple of exponential patterns, and ending up with a current logarithmic pattern revealing that the evolution of the Greek COVID-19 infection-curve tends into saturation. The network analysis also illustrates stability of hubs and instability of medium and low-degree nodes, implying a low probability to meet maximum (infection) values at the future and high uncertainty in the variability of other values below the average. The overall approach contributes to the scientific research by proposing a novel method for the structural decomposition of a time-series into periods, which allows removing from the series the disconnected past-data facilitating better forecasting, and provides insights of good policy and decision-making practices and management that may help other countries improve their performance in the war against COVID-19.

**Keywords:** Coronavirus; pandemics; infectious disease; natural visibility algorithm; community detection; modularity optimization.

## 1. INTRODUCTION

Starting at December 2019, in Wuhan city (China), the COVID-19 coronavirus disease is a new epidemic being virally spread across the world, causing cascading deaths and uncertainty for the future of the global and national economies (Anderson et al., 2020; McKibbin et al., 2020; WHO, 2020; Wu et al., 2020). Prime epidemiological studies on the COVID-19 infection and death curves (Li et al., 2020a,b; Liu et al., 2020) reveal the existence of strong scaling dynamics in the spreading of the disease, which appears to initiate under an exponential growth. While the major concern of the research community and academia is obviously to find a cure for defeating this lethal pandemic (Gao et al., 2020; Gautret et al., 2020), it is a consensus that (even when a laboratory formula will be soon available) the process of moving a cure from the laboratory to the production phase cannot be done at once. Therefore, beyond the global expectations that the day of a cure to this pandemic is forthcoming, governments worldwide are currently facing the challenge of managing their medical and human resources against the scaling (exponential-alike) trends of the disease-spreading (Anderson et al., 2020; McKibbin et al., 2020; Remuzzi and Remuzzi, 2020). The case of Italy was among the first countries dealing with COVID-19 outside the country of its origin, China. The uncontrolled spreading of the disease and the number of casualties (Roser and Ritchie, 2020) in Italy imply, on the one hand, the inertia of social life and structures to abrupt changes (Liningston and Bucher, 2020) and, on the other hand, the small-world (Watts and Strogatz, 1998; Christakis and Fowler, 2009; Tsiotas, 2019) configuration of modern societies developing network structures. Although current scientific knowledge have led to the development of standardized protocols, best practices, and policies for controlling the pandemic (Bedford et al., 2020; Sohrabi, 2020), countries worldwide show diverse patterns of conformance to these instructions, determining thus the evolution of the disease in their population in different ways. For instance, USA, UK, and Sweden, which appear to build up (at least initially) to the "herd immunity" strategy (Cohen, 2020) are currently included in the top-ten countries in COVID-19 mortality rates (Roser and Ritchie, 2020), whereas others showing early emergence seem to succeed suspending their casualties.

Greece is an insightful case of good performance for its anti-COVID-19 policies, which currently manages to keep the infected cases and deaths at relatively low levels (Roser and Ritchie, 2020; Xu et al., 2020). In particular, Greece faced the first case of infection on February 26$^{th}$, 2020, and just three days later it started applying several policies for controlling the disease. This timely alert and response have led Greece to be currently in the last places of the European Countries, in terms of infection and mortality rates of COVID-19 (Roser and Ritchie, 2020; Xu et al., 2020), setting the conditions to consider its case as a success story in the war against the pandemic. Within the context of promising a success story, this paper studies the evolution of the COVID-19 infection-curve in Greece in relation to the anti-COVID-19 policies applied by the state to control the disease aiming to detect whether and how these policies affected the dynamics of the disease spreading. Provided that the spreading of COVID-19 in Greece counts a less than two months period, the classic time-series analysis approaches are inevitably subjected to high level of uncertainty for forecasting the future of the epidemic (Box et al., 2015). Besides, prime quantitative approaches have already sufficed to reveal the exponential dynamics in the spreading of the disease (Li et al., 2020a,b; Liu et al., 2020). Therefore, this paper advances current quantitative approaches by building on complex network analysis of time-series to transform the COVID-19 infection-curve to a complex network. To do so, the visibility graph algorithm is applied (Lacasa et al., 2008), which is a non-parametric approach able to provide structural insights of the time-series and to transform it to a complex network that is a tensor of higher dimension in comparison to the series (Box et al., 2015; Tsiotas and Charakopoulos, 2020). The associated to the time-series visibility graph is detected for communities, based on the modularity optimization algorithm (Blondel et al., 2008), aiming to develop a partition of the time-series and therefore to interpret their significance in accordance with the anti-COVID-19 policies applied by the Greek state.

The overall approach, first, aims to provide insights of good policy and decision making practices and management that may help other countries improve their performance against the pandemic. Secondly, it contributes to the scientific research by proposing a novel method for the structural decomposition of a time-series into periods that allow removing from the series the disconnected past data and thus it facilitates better forecasting. Both the proposed methodological framework and the cases of the current success story of Greece can be proven effective in the war against COVID-19.

**2. METHODOLOGY AND DATA**
The methodological framework builds on complex network-based analysis of time-series to transform the time-series of the COVID-19 infection-curve in Greece to a complex network and then to examine the associated complex network instead of the time-series. The methodology consists of three steps; at the first, an associated visibility graph is generated from the Greek COVID-19 infection-curve where its degree distribution is examined. At the second step, the associated visibility graph is divided into communities according to the modularity optimization algorithm (Blondel et al., 2008) and then the time-series corresponding to each community are examined by using curve fitting techniques. At the final step, the Greek anti-COVID-19 policies included in each community are discussed and a policy profile for each community is attempted to develop. The dataset considered in the analysis includes the starting day (February 26[th], 2020) where the first infection emerged in Greece, up to the 43[rd] day (April 4[th], 2020). Each day following the last day included in the time-series (up to the publication of the preprint of this paper) is used to verify forecasting ability of each community.

**2.1. Generating a visibility graph**
Lacasa et al. (2008) proposed an algorithm, the natural visibility algorithm (NVG), for transforming a time-series into a graph (complex network) $G_V(V,E)$. The NVG builds on the intuition that each node of a time series can be seen as a mountain with an observer standing on it, where other nodes are visible to each observer according to their heights (values). In technical terms, each node $(t_k, x(t_k))$ of the time-series corresponds to a graph node $n \in V$ (preserving the time-series order) and two nodes $n_i, n_j \in V$ are connected $(n_i, n_j) \in E$ when a visibility line between the corresponding time-series data can be drawn (i.e. if it is possible to draw a straight line between values without interception with other values), as it is shown in Fig.1. Within the "landscape" configured by the time-series, a node (bar) is visible by a node ($n_o$) when no other node intersects the visibility lines originating from $n_o$.

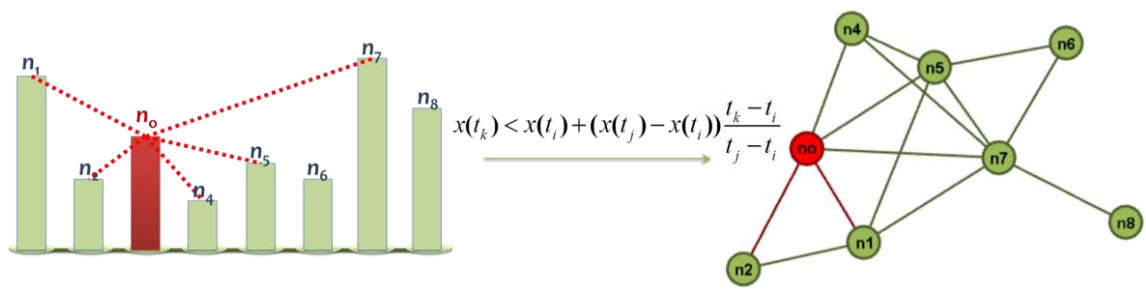

**Fig.1.** Example of the natural visibility algorithm (NVG), which generates a network (right side) from a time-series (left side). Dashed lines in the time-series' plot represent the visibility lines.

The mathematical expression of the NVG connectivity criterion is given by inequality:

$$x(t_k) < x(t_i) + (x(t_j) - x(t_i))\frac{t_k - t_i}{t_j - t_i} \qquad (1),$$

where $x(t_i)$ and $x(t_j)$ express the numeric values of the time-series nodes and $t_i$, $t_j$ their time-reference points. In a network mapped according to the visibility algorithm, each node is visible at least by its nearest (left and right) neighbors (Gao et al., 2017; Tsiotas and Charakopoulos, 2020).

## 2.2. Community detection based on modularity optimization

The associated to the Greek COVID-19 infection-curve visibility graph $G_V(V,E)$, which is generated by the NVG algorithm, is divided to communities by using the modularity optimization algorithm of Blondel et al. (2008). This algorithm produces a number of communities under the criterion to maximize the intra-community (i.e. within the communities) connectivity and therefore to minimize the inter-community (between the communities) connectivity, as denoted by the expression:

$$\text{maximize } [Q \propto (m_{within\ communities} - m_{between\ communities})] \qquad (2),$$

where $Q$ is the modularity function (Fortunato, 2010) and $m$ the number of links. The modularity optimization algorithm (Blondel et al., 2008) is a greedy approach applied in two steps. At the first step, all graph nodes are assigned into different communities. Next, the nodes are sequentially being swept and placed to collective communities if the assignment of a node into a neighbor community increases the gain in the weighted modularity function ($Q_w$) of the initial graph. At the second step, the collective communities are replaced by super-nodes and the procedure is repeated until the modularity function cannot increase any more (Fortunato, 2010; Tsiotas, 2019).

## 3. RESULTS AND DISCUSSION
### 3.1. Network analysis

The degree distribution $p(k)$ of the associated to the Greek COVID-19 infection-curve visibility graph ($G_V$) is shown at Fig.2a. Among a set of available fittings (exponential, Fourier, Gaussian, linear, up to $4^{th}$ order polynomial, power, and logarithmic) that were tested (Chapra, 2012) to the data, the normal (Gaussian) curve was found to have both the highest coefficient of determination ($R^2$=0.761) and the simplest expression (least number of estimated terms). Comparatively to a power-law curve, which is a fitting of special importance in network science for the detection of the scale-free property (Tsiotas, 2019) and thus of hierarchical structures, the normal (Gaussian) curve with an estimated average $\hat{\mu} = \langle \hat{k} \rangle$ =4.491 (number of connections per node) and an standard deviation $\hat{\sigma}$ =2.518 better fits to the $p(k)$ data. The normal-shaped pattern of the degree distribution may have a double interpretation. On the one hand, it implies that the network is ruled by randomness, provided that a normal curve asymptotically approximates the binomial distribution, which theoretically describes the degree distribution of random network (Barabasi, 2016). On the other hand, it expresses that the network is subjected to spatial constraints (Barthelemy, 2016) and therefore that the associated visibility graph has more a mesh-alike rather than a hierarchical structure. To get more insights about the topology of the $G_V$, we examine the sparsity (spy) plot of this visibility graph (Fig.2b), which is a plot of the adjacency matrix displaying nonzero elements with dots (Tsiotas, 2019). The major concentration of the nonzero elements towards the main diagonal of the matrix implies the existence of spatial constraints (Tsiotas, 2019) in the visibility network, which verifies the previous observation based on the degree distribution consideration. However, the vertical spreading of dots in the spy plot may also indicate random trends (Tsiotas, 2019), which is also in line with the previous observations.

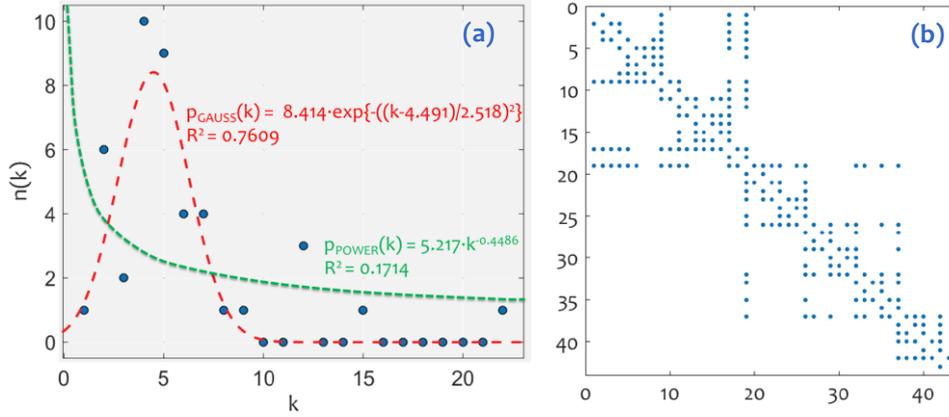

**Fig.2.** (a) The degree distribution $p(k)$ of the associated to the Greek COVID-19 infection-curve visibility graph $G_V$ ($n(k)$ denotes the frequency of degree $k$), (b) The sparsity (spy) plot of $G_V$.

According to the degree distribution and sparsity pattern consideration, the Greek COVID-19 visibility graph is configured by random forces and spatial constraints. The emergence of randomness illustrates that future connections (which are created based on the daily infection values) may randomly attach to the network and therefore the future maximum values of daily infection are possible. On the other hand, indications of spatial constraints imply that future connections of the network are more depended on their neighborhood values (i.e. the previous states) and thus the phenomenon of the disease can resemble to a Markovian chain. However, regardless the insignificant level of power-law determination ($R^2$=0.174), a long-tail behavior of the degree distribution can be observed for degree values higher than 10 ($k > 10$), which expresses that the network structure of the $G_V$ is steadier for the hubs (i.e. nodes of high degree) than this of medium and low-degree nodes. This observation implies that the probability to meet maximum (infection) values (and therefore new hubs in the visibility network) at the future is considerably low, in contrast to the medium and low-degree nodes, where their future behavior is more uncertain.

### 3.2. Community detection
▪ *Structural analysis*

At the second step, the associated to the Greek COVID-19 infection-curve visibility graph ($G_V$) is divided into communities, on which connectivity within each community is dense and between communities is sparse. The community discrimination of $G_V$ is shown in Fig.3, where labels of network nodes express the day since the first infected case.

The community detection analysis of $G_V$ resulted to five discrete communities (Fig.3b:g), which appear into a sequential ordering ($Q_1$, $Q_2$, $Q_1$, $Q_3$, $Q_4$, $Q_5$), except the second community ($Q_2$) that is injected (intermediated) to the first one. This mixed configuration of the first and second community ($Q_1$, $Q_2$, $Q_1$,...) suggests a novel finding due to modularity optimization approach, which cannot (or hardly can) be detected by other established time-series approaches. As being evident in Fig.3, the evolution of the Greek COVID-19 infection-curve is vertebral and consists of five discrete time-periods (parts or communities). The first ($Q_1$) starts from the $1^{rst}$ day and ends up to the $19^{th}$ day (Fig.3c) of the infection spreading in Greece and is described by a scaling (power) pattern ($R_1^2$=0.98) with an exponent $a_1 \approx 2.21$. The second part ($Q_2$) is injected in the first one and is extended to the time-interval $Q_2$=[5,8]={5,6,7,8} $\subset \mathbb{N}$ (Fig.3d), described by a ($2^{nd}$ order) polynomial (U-shaped) pattern ($R_2^2$=0.933) with a positive coefficient ($a_2 \approx 0.5$) of the highest order term. The third part ($Q_3$) follows the first one, is extended to the time-period $Q_3=[20,26]\subset \mathbb{N}$ (Fig.3e), and is described by an exponential pattern ($R_1^2=0.986$) of exponent $a_3\approx 0.09$. Next, the fourth part ($Q_4$) is extended to the time-interval $Q_4=[27,32]\subset \mathbb{N}$ (Fig.3f) and it is also described by an exponential pattern ($R_1^2=0.986$), but of smaller exponent than the previous period ($a_4\approx 0.085 < a_3$). Finally, the fifth part ($Q_5$) is extended to the time-interval $Q_4=[33,43]\subset \mathbb{N}$ (Fig.3f) and is described by a logarithmic pattern ($R_1^2=0.983$).

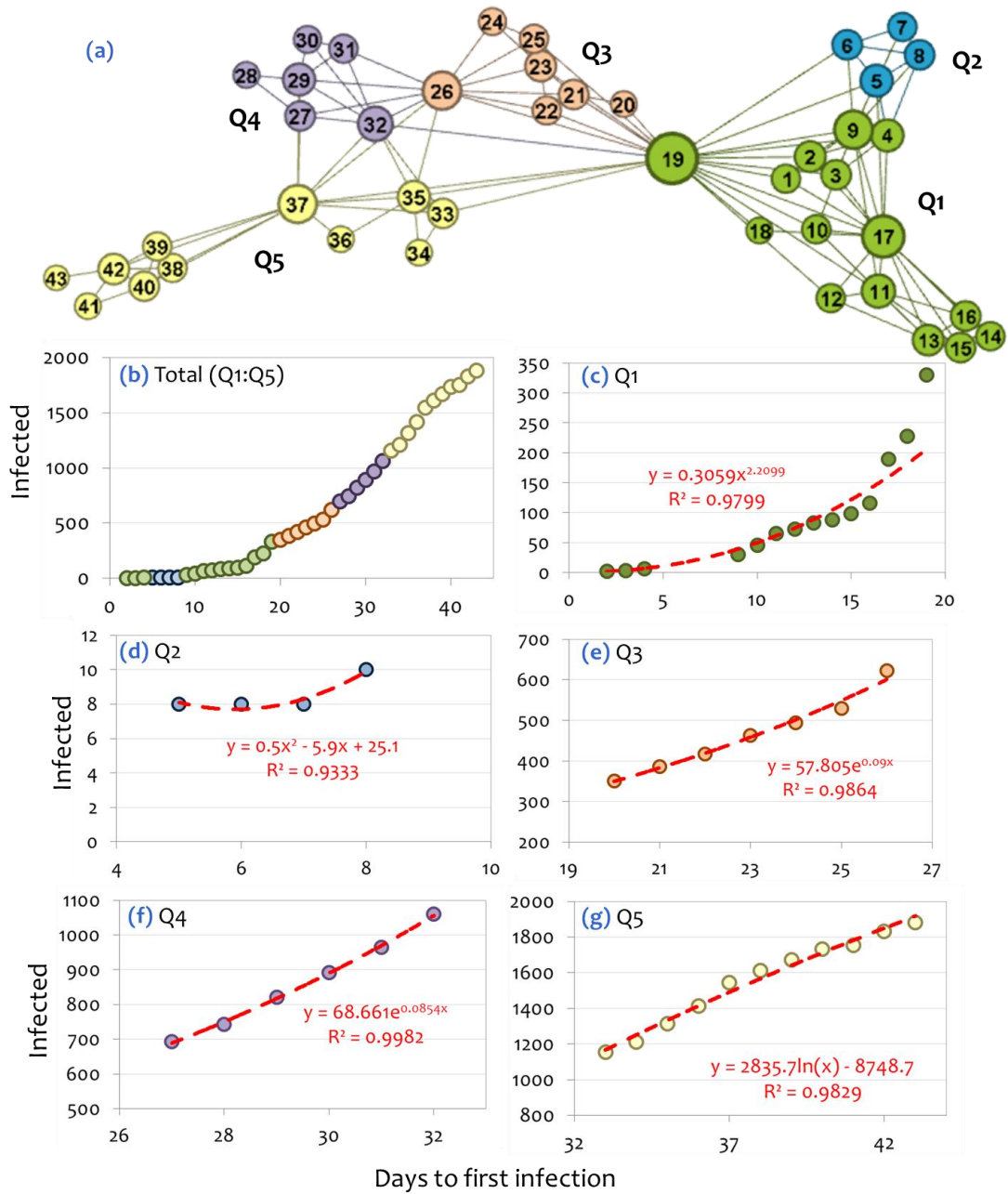

**Fig.3.** (a) Division of the associated to the Greek COVID-19 infection-curve visibility graph ($G_V$) into five communities ($Q_1:Q_5$) based on the modularity optimization algorithm of Blondel et al. (2008), (b) the community structure of the Greek COVID-19 infection-curve, where (c) up to (g) show best fittings (of the highest $R^2$) within each community.

Overall, according to the sequence of the modularity communities in the time-series body, the Greek COVID-19 infection-curve initiated the first three days ($Q_1$) with scaling (power) dynamics. Next, it showed an abrupt (U-shaped) tendency at the next four days ($Q_2$), which was returned to the previous scaling (power) status ($Q_1$), where it remained for the next 11 days [9,19]⊂ℕ. The next 13 days [20,32]⊂ℕ, the Greek COVID-19 infection-curve showed an exponential growth but in the mid-time of the period the exponent was decreased, and finally the last 11 days ([33,43]⊂ℕ) the infection-curve showed a logarithmic trend implying that the evolution of the infection-curve leads to saturation.

▪ *Forecasting*

Based on the fittings applied to each community ($Q_1$:$Q_5$), future infections of the COVID-19 infection-curve for the next (to the 43rd day) 11 days are estimated and shown in Table 1. In this paper, the analysis is based on the Greek COVID-19 infection time-series, starting from the 1st day and ending at the 43rd day since the first infection (namely on the set [1,43]⊂ℕ). Post to the 43rd day cases (emerging at the time writing this paper) are registered in Table 1 to test the estimations (as verification set).

**Table 1**
Estimation of future cases of the COVID-19 infection-curve based on the parametric fittings applied to each community ($Q_1$:$Q_5$).

| Date | n | Observed Infections[*] | $Q_1$ | Diff | $Q_2$ | Diff | $Q_3$ | Diff | $Q_4$ | Diff | $Q_5$ | Diff |
|---|---|---|---|---|---|---|---|---|---|---|---|---|
| 4/9/20 | 44 | 1955 | 1311 | 644 | 734 | 1222 | 3032 | -1077 | 2942 | -987 | 1982 | -27 |
| 4/10/20 | 45 | 2011 | 1377 | 634 | 772 | 1239 | 3318 | -1307 | 3204 | -1193 | 2046 | -35 |
| 4/11/20 | 46 | 2081 | 1446 | 635 | 812 | 1269 | 3630 | -1549 | 3490 | -1409 | 2108 | -27 |
| 4/12/20 | 47 | pending[*] | 1516 | pending | 852 | pending | 3972 | pending | 3801 | pending | 2169 | pending |
| 4/13/20 | 48 | pending | 1588 | pending | 894 | pending | 4346 | pending | 4140 | pending | 2229 | pending |
| 4/14/20 | 49 | pending | 1662 | pending | 937 | pending | 4756 | pending | 4509 | pending | 2287 | pending |
| 4/15/20 | 50 | pending | 1738 | pending | 980 | pending | 5203 | pending | 4911 | pending | 2345 | pending |
| 4/16/20 | 51 | pending | 1816 | pending | 1025 | pending | 5693 | pending | 5349 | pending | 2401 | pending |
| 4/17/20 | 52 | pending | 1896 | pending | 1070 | pending | 6230 | pending | 5825 | pending | 2456 | pending |
| 4/18/20 | 53 | pending | 1977 | pending | 1117 | pending | 6816 | pending | 6345 | pending | 2510 | pending |
| 4/19/20 | 54 | pending | 2061 | pending | 1165 | pending | 7458 | pending | 6910 | pending | 2563 | pending |

*. Due to the evolving status of the COVID-19 infection-curve, the overall analysis was applied up to the 43rd day since the first infection. Additional cases appearing at the time preparing the paper are entered in the column named "observed infections" and are used as verification set to the estimations shown in columns corresponding to the $Q_1$:$Q_5$ communities.

As it can be observed in Table 1, the logarithmic curve of the last modularity community interval ($Q_5$) provides the most accurate estimations of the three updated days amongst the five available community intervals. Moreover, differences of most of communities seem to be of a certain magnitude (~600 for $Q_1$, ~1200 for $Q_2$, ~ −1200 for $Q_3$, ~ −1100 for $Q_4$, and ~ −30 for $Q_5$), which may be related to the effectiveness of the overall community detection approach.

### 3.3. Policy assessment

At the final step, the communities resulted from the modularity optimization algorithm are examined in accordance with the anti-COVID-19 policies applied by the Greek state in the meantime. Table A1 (shown in the appendix) summarizes the evolution of the major policies of the state in the fight against the disease, which are grouped according to the communities resulted by the analysis. As it can be observed, the first three days ($\in Q_1$) of evolution of the Greek COVID-19 infection-curve, the state applied optional policy measures for the control of the disease. The first compulsory measure was taken at the end of the (intermediated) $Q_2$ period, applied at local scale, showing perhaps an alerted of the state for the changing dynamics of the disease. During the second half of the first ($Q_1$) period, which is placed between the 9th to 19th days since the first infection, a set of compulsory measures was taken for the first time at the national scale, concerning suspension of educational, cultural, justice, and recreation functions and activities. At the next two periods ($Q_3$ and $Q_4$) of exponential growth, more severe measures were applied. Measures of $Q_3$ period concern national suspension of gathering, religious, and tourism activities and of maritime transport. On the other hand, the defining measure of the $Q_4$ period was started at its beginning and concerns the national restriction of road transport. The defining measure of the final period ($Q_5$) regards the control of transportation and particularly the national suspension of private air transport. Overall, the time evolution of the major Greek anti-COVID-19 policies in comparison to the structure of the modularity optimization time-periods of the Greek COVID-19 infection-curve, signify a timely alert of the Greek state to the cascading dynamics of the pandemic. The evolution in the change of type of the fitting curve, from power to polynomial, to power, to exponential, to a smoother exponential, and currently (at the time writing this paper) to logarithmic reveal that anti-COVID-19 policies applied by the Greek state succeed currently to control the cascading spread of the disease.

## 4. CONCLUSIONS

The Greek anti-COVID-19 policies were applied just three days after the emergence of the first infected case in the country appear determinative to the evolution of the COVID-19 infection-curve. The study built on complex network analysis of time-series (based on the visibility graph algorithm) to detect dynamics in the structure of the infection-curve that are not visible by using classic time-series analysis methods, due to data limitation. Network analysis applied to the time series showed the existence of spatial constraints mixed with random trends, illustrating that (in the long-term) the evolution of the Greek COVID-19 infection-curve resembles to a Markovian chain with slightly random transitions. In terms of forecasting, the structural analysis of the visibility graph detected stability of the hubs and instability of the medium and low-degree nodes, implying a low probability to meet maximum (infection) values at the future and high uncertainty in the variability of medium and lower values. Based on the modularity optimization algorithm, the Greek COVID-19 associated visibility graph is divided to five stages (periods) of different connectivity ruling the evolution of the pandemic in Greece. The first stage is described by a power pattern, the second (which intermediates the first one) is described by a second order polynomial (U-shaped) pattern, the third and fourth by exponential patterns, and a final one by a logarithmic pattern implying that the Greek COVID-19 infection-curve tends to saturate. The resulting of the Greek COVID-19 infection-curve to a logarithmic pattern can illustrate the effectiveness of the anti-COVID-19 policies applied in Greece, during the 43 days since the emergence of the first infected case. This measurable evolution COVID-19 infection-curve allows currently considering Greece as a success story in the war against the pandemic. However, a long way is still pending and in conjunction with the global expectations that the day of a cure to this pandemic is forthcoming, this paper provides insights for the current success story of Greece to inspire other countries towards more effective anti-COVID-19 policies and management.

## REFERENCES
Anderson, R. M., Heesterbeek, H., Klinkenberg, D., & Hollingsworth, T. D. (2020). How will country-based mitigation measures influence the course of the COVID-19 epidemic?. *The Lancet*, *395*(10228), 931-934.
Barabasi, A. L., (2016) *Network Science*, Cambridge, UK, Cambridge University Press.

Barthelemy, M., (2011) "Spatial networks", Physics Reports, 499, pp.1–101.

Bedford, J., Enria, D., Giesecke, J., Heymann, D. L., Ihekweazu, C., Kobinger, G., ... & Ungchusak, K. (2020). COVID-19: towards controlling of a pandemic. *The Lancet*.

Blondel, V., Guillaume, J.-L., Lambiotte, R., Lefebvre, E., (2008) "Fast unfolding of communities in large networks", *Journal of Statistical Mechanics*, 10, P10008. https://doi.org/10.1088/1742-5468/2008/10/P10008.

Box, G., Jenkins, G. M., Reinsel, G. C., Ljung, G. M. (2015) *Time series analysis: forecasting and control*, John Wiley & Sons, New Jersey.

Chapra, S. C. (2012). *Applied numerical methods with MATLAB for engineers and scientists*. New York: McGraw-Hill.

Christakis, N. A., & Fowler, J. H. (2009). *Connected: The surprising power of our social networks and how they shape our lives*. Little, Brown Spark.

Cohen, J., & Kupferschmidt, K. (2020). Countries test tactics in 'war' against COVID-19.

Fortunato, S., (2010) "Community detection in graphs", Physics Reports, 486, pp.75–174. https://doi.org/10.1016/j.physrep.2009.11.002.

Gao Z-K., Small M., Kurths J., (2017) "Complex network analysis of time-series", Europhysics Letters;116(5), 50001.

Gao, J., Tian, Z., & Yang, X. (2020). Breakthrough: Chloroquine phosphate has shown apparent efficacy in treatment of COVID-19 associated pneumonia in clinical studies. *Bioscience trends*.

Gautret, P., Lagier, J. C., Parola, P., Meddeb, L., Mailhe, M., Doudier, B., ... & Honoré, S. (2020). Hydroxychloroquine and azithromycin as a treatment of COVID-19: results of an open-label non-randomized clinical trial. *International Journal of Antimicrobial Agents*, 105949.

Lacasa, L., Luque, B. , Ballesteros, F., Luque, J., Nuno, J.C., (2008) "From time-series to complex networks: The visibility graph", *Proceedings of the National Academy of Sciences*, 105(13), pp.4972–4975.

Li, M., Chen, J., & Deng, Y. (2020). Scaling features in the spreading of COVID-19. *arXiv preprint arXiv:2002.09199*.

Li, Y., Liang, M., Yin, X., Liu, X., Hao, M., Hu, Z., ... & Jin, L. (2020). COVID-19 epidemic outside China: 34 founders and exponential growth. *medRxiv*.

Liu, Y., Gayle, A. A., Wilder-Smith, A., & Rocklöv, J. (2020). The reproductive number of COVID-19 is higher compared to SARS coronavirus. *Journal of travel medicine*.

Livingston, E., & Bucher, K. (2020). Coronavirus disease 2019 (COVID-19) in Italy. *Jama*.

Mahase, E. (2020). Covid-19: UK starts social distancing after new model points to 260 000 potential deaths.

McKibbin, W. J., & Fernando, R. (2020). The global macroeconomic impacts of COVID-19: Seven scenarios.

Remuzzi, A., & Remuzzi, G. (2020). COVID-19 and Italy: what next?. *The Lancet*.

Roser, M., Ritchie, H., (2020) "Coronavirus Disease (COVID-19)" available at the URL: https://ourworldindata.org/coronavirus-data [accessed: 10/4/20]

Sohrabi, C., Alsafi, Z., O'Neill, N., Khan, M., Kerwan, A., Al-Jabir, A., ... & Agha, R. (2020). World Health Organization declares global emergency: A review of the 2019 novel coronavirus (COVID-19). *International Journal of Surgery*.

Tsiotas, D., (2019) "Detecting different topologies immanent in scalefree networks with the same degree distribution", *Proceedings of the National Academy of Sciences*, 116(14), pp.6701–6.

Tsiotas, D., Charakopoulos, A., (2020) "VisExpA: Visibility expansion algorithm in the topology of complex networks", *Software X*, 11, 100379.

Watts, D., Strogatz, D., (1998) "Collective dynamics of small-world networks", *Nature*, 393, pp.440–442.
World Health Organization. (2020). Coronavirus disease 2019 (COVID-19): situation report, 72.
Wu, J. T., Leung, K., Bushman, M., Kishore, N., Niehus, R., de Salazar, P. M., ... , Leung, G. M., (2020) "Estimating clinical severity of COVID-19 from the transmission dynamics in Wuhan, China", *Nature Medicine*, pp.1-5.
Xu, B., Gutierrez, B., Mekaru, S., Sewalk, K., Goodwin, L., Loskill, A., ... , Zarebski, A. E., (2020) "Epidemiological data from the COVID-19 outbreak, real-time case information", *Scientific Data*, *7*(1), pp.1-6.

**APPENDIX**

**Table A1**

Time evolution of major Greek anti-COVID-19 policies[*].

| Date | Community | n | Daily Infections | Total Infections | Policy Measure |
|---|---|---|---|---|---|
| 2/27/20 | #1 | 2 | 2 | 2 | |
| 2/28/20 | | 3 | 1 | 3 | Optional local suspension of schools operation at regions with infected cases; Prohibition of school excursions abroad, school decontaminations. |
| 2/29/20 | | 4 | 3 | 6 | |
| 3/1/20 | #2 | 5 | 2 | 8 | |
| 3/2/20 | | 6 | 0 | 8 | |
| 3/3/20 | | 7 | 0 | 8 | |
| 3/4/20 | | 8 | 2 | 10 | Local suspension of schools, theaters, cinemas, and cultural spaces at regions with infected cases. |
| 3/5/20 | #1 | 9 | 20 | 30 | |
| 3/6/20 | | 10 | 15 | 45 | |
| 3/7/20 | | 11 | 20 | 65 | |
| 3/8/20 | | 12 | 7 | 72 | National suspension of nursing houses, conferences, and school excursions and indoors operation of sport activities. |
| 3/9/20 | | 13 | 11 | 83 | National suspension of cultural and art activities. |
| 3/10/20 | | 14 | 5 | 88 | National suspension of schools and educational units. |
| 3/11/20 | | 15 | 10 | 98 | |
| 3/12/20 | | 16 | 18 | 116 | National suspension of playgrounds, theaters, cinemas, art and entertainment places. |
| 3/13/20 | | 17 | 73 | 189 | National suspension of courts, malls, café, bars, restaurants, museums, sport centers and organized beaches. |
| 3/14/20 | | 18 | 38 | 227 | |
| 3/15/20 | | 19 | 103 | 330 | |
| 3/16/20 | #2 | 20 | 21 | 351 | National establishment of money penalty for disobedience to the measures and of compulsory 14-day quarantine to those entering the country. |
| 3/17/20 | | 21 | 35 | 386 | National suspension of religion activities, closing of national borders for tourism-related purposes. |
| 3/18/20 | | 22 | 31 | 417 | National prohibition of public gathering (>10 persons), regulations to supermarket's entrance. |
| 3/19/20 | | 23 | 46 | 463 | National suspension of hotels. |
| 3/20/20 | | 24 | 31 | 494 | National restriction to maritime transport. |
| 3/21/20 | | 25 | 35 | 529 | |
| 3/22/20 | | 26 | 94 | 623 | |
| 3/23/20 | #3 | 27 | 71 | 694 | National restriction of road transport. |
| 3/24/20 | | 28 | 48 | 742 | |
| 3/25/20 | | 29 | 78 | 820 | |
| 3/26/20 | | 30 | 71 | 891 | |
| 3/27/20 | | 31 | 74 | 965 | |
| 3/28/20 | | 32 | 95 | 1060 | |
| 3/29/20 | #4 | 33 | 95 | 1155 | |
| 3/30/20 | | 34 | 56 | 1211 | |
| 3/31/20 | | 35 | 102 | 1313 | |
| 4/1/20 | | 36 | 101 | 1414 | National suspension of private air transport. |
| 4/2/20 | | 37 | 129 | 1543 | |
| 4/3/20 | | 38 | 69 | 1612 | |
| 4/4/20 | | 39 | 60 | 1672 | |
| 4/5/20 | | 40 | 62 | 1734 | |

| Date | | | | |
|---|---|---|---|---|
| 4/6/20 | | 41 | 20 | 1754 |
| 4/7/20 | | 42 | 77 | 1831 |
| 4/8/20 | | 43 | 52 | 1883 |

Source: Own elaboration based on government discourses and newspapers